\documentclass{iaus}
\usepackage{graphicx}
\usepackage{natbib}

\title[Profiling Young Massive Stars] 
{Profiling Young Massive Stars}

\author[Hill, Burton, Cunningham \& Minier ]   
{Tracey Hill$^{1,2}$%
  \thanks{Present address: Leiden Observatory, Leiden University, PO BOX 9513, 2300RA Leiden, the Netherlands.},
 M. G. Burton$^2$ \break M. R. Cunningham$^2$ \and V. Minier$^3$}

\affiliation{$^1$ Leiden Observatory, Leiden University, PO BOX 9513, 2300RA, Leiden, the Netherlands. \break email: thill@strw.leidenuniv.nl \\[\affilskip]
$^2$ School of Physics, University of New South Wales, Kensington, NSW 2052, Australia\\[\affilskip]
$^3$Service d'Astrophysique, DAPNIA/DSM/CEA Saclay, 91191 Gif-sur-Yvette, France.}

\pubyear{2007}
\volume{242}  
\pagerange{119--126}
\date{?? and in revised form ??}
\jname{Proceedings Title IAU Symposium}
\editors{A.C. Editor, B.D. Editor \& C.E. Editor, eds.}

\newcommand{\hii}{{H{\scriptsize II}}}
\newcommand{\uchii}{{UC H{\scriptsize II}}}
\newcommand{\IRAS}{{\it IRAS}}

\newcommand{\MSX}{{\it MSX}}
\newcommand{\microm}{{$\mu$m}}

\begin{document}

\maketitle

\begin{abstract}
   We present the results of spectral energy distribution analysis for 162 of the 405 sources reported in the SIMBA survey of \cite{hill05}. The fits reveal source specific parameters including: the luminosity, mass, temperature, H$_2$ number density, the surface density and the luminosity-to-mass ratio. Each of these parameters are examined with respect to the four classes of source present in the sample. Obvious luminosity and temperature distinctions exist between the mm-only cores and those cores with methanol maser and/or radio continuum emission, with the former cooler and less luminous than the latter. The evidence suggests that the mm-only cores are a precursor to the methanol maser in the formation of massive stars. The mm-only cores comprise two distinct populations distinguished by temperature. Analysis and conclusions about the nature of the cool-mm and warm-mm cores comprising the mm-only population are drawn.

\keywords{masers, stars: formation, stars: fundamental parameters, stars: early-type, radio continuum: stars, methods: data analysis}
\end{abstract}

\firstsection 
\section{Introduction}

   Despite the wealth of attention that massive stars have received of late, their formation scenario is still largely unclear. This may be attributed to the fact that massive stars form in clustered mode, deeply embedded in their natal molecular envelope where their development is optically obscured prior to main sequence evolution. This, coupled with the rapid timescales over which they evolve, as well as the fact that high-mass stellar candidates are typically at further distances than their lower mass counterparts, hinders the study of these objects.
  
   Massive star formation regions are often found coincident with maser sources -- in particular the methanol variant \citep{pestalozzi05}, \uchii\, regions \citep{thompson06}, \IRAS\, colour selected sources \citep{wood89morph} and \MSX\, colour selected sources \citep{lumsden02}. More specifically, methanol masers and \uchii\, regions are thought to feature prominently in the earliest stages of massive star formation \citep[cf.][]{batrla87, caswell95, walsh97, minier01, beuther02, faundez04, thompson06}.

 Methanol masers have been found associated with strong radio continuum emission (i.e. \hii\, regions), \IRAS\, far-infrared colour selected sources \citep[e.g.][]{wood89col}, H$_{2}$O and OH masers \citep{caswell95} as well as millimetre and submillimetre continuum emission \citep[][respectively]{hill05, walsh03}.

 Early work \citep[e.g.][]{walsh98} suggested that the methanol maser was indicative of the earliest stages of massive star formation prior to the onset of radio continuum emission. More recent work has focused on finding a precursor to the methanol maser, which would mark the {\it very} earliest stages of massive star formation. The hot molecular core (HMC) is one such object proposed to satisfy this criterion \citep[cf.][]{olmi96, osorio99, minier05}.

   \citet[][hereafter, Paper I]{hill05} undertook a millimetre continuum emission survey toward regions displaying evidence of massive star formation in search of cold cores that would mark the earliest stages of their formation. This survey revealed evidence of star formation clearly offset from, and devoid of, both the methanol maser and radio continuum sources targeted. Follow-up submillimetre observations of these `mm-only' cores \citep[][hereafter Paper II]{hill06} revealed each of them to be associated with submillimetre continuum emission and hence cold deeply embedded objects.

\section{Spectral Energy Distribution Fitting}\label{sed}

  Combining the SIMBA and SCUBA data from Papers I and II, together with existing submillimetre SCUBA data \citep{p-p00, walsh03, thompson06}, as well as archival \MSX\, data where applicable and in some instances \IRAS, we have drawn spectral energy distribution (SED) diagrams for the sources in our sample. The SED is drawn using a two-component Levenberg-Marquardt least squares fit according to equation \ref{chap:sed:eqfit}. This relation models emission from a hot core embedded in a larger cold or warm envelope \citep[cf.][]{minier05}.

\vspace{-0.1cm}
 \begin{equation}\label{chap:sed:eqfit}
F_\nu = [\Omega_h B_\nu(T_{hot})] + \Omega_c B_\nu(T_{cold})\, \epsilon_\nu
\end{equation}

\noindent
where $F_\nu$ is the flux density of the source, $\Omega_h$ and $\Omega_c$ are the source solid angles for the hot and the cold component of the source, respectively, $B_\nu$ is the Planck function for a temperature of $T_{hot}$ and $T_{cold}$, and $\epsilon_{\nu}$ is the emissivity for the cold component, which is equivalent to (1\,-\,e$^{-\tau_{\nu}}$). For optically thin regions, $\epsilon_{\nu}$ approximates to $\tau_{\nu}$ where $\tau_{\nu}$ is given by $\tau_0(\nu/\nu_0)^{\beta}$. The dust grain emissivity index ($\beta$) is assumed to be 2 for the sample, as per Paper II.

   SED fits were applied to 162 of the 405 sources in the SIMBA sample (see Fig. 1).

\begin{figure}
\begin{center}
 \includegraphics[height=6.4cm, width=8.5cm]{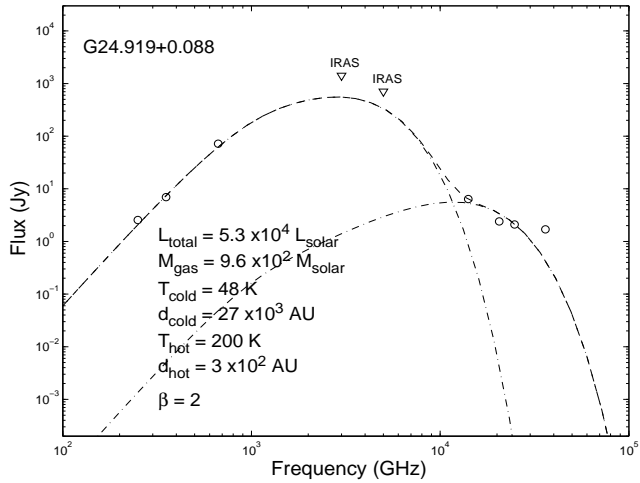}
  \caption{Spectral energy distribution of the mm-only core G24.919+0.088. \IRAS\, 60 and 100\microm\, upper limits are indicated on the diagram. Source specific parameters are as listed.}
\label{fig:sed}
\end{center}
\end{figure}

\section{Results and Analysis}\label{results}

   The SED fit provided three main parameters: the dust temperature, luminosity and the mass. These were then used to determine the hydrogen number density (n$_{H_2}$), the surface density ($\Sigma$) and the luminosity-to-mass ratio (L/M). Together with the radius and distance from our earlier work, eight parameters are known for each source.

\begin{figure}
  \begin{center}
    \includegraphics[height=5.5cm, width=6.7cm]{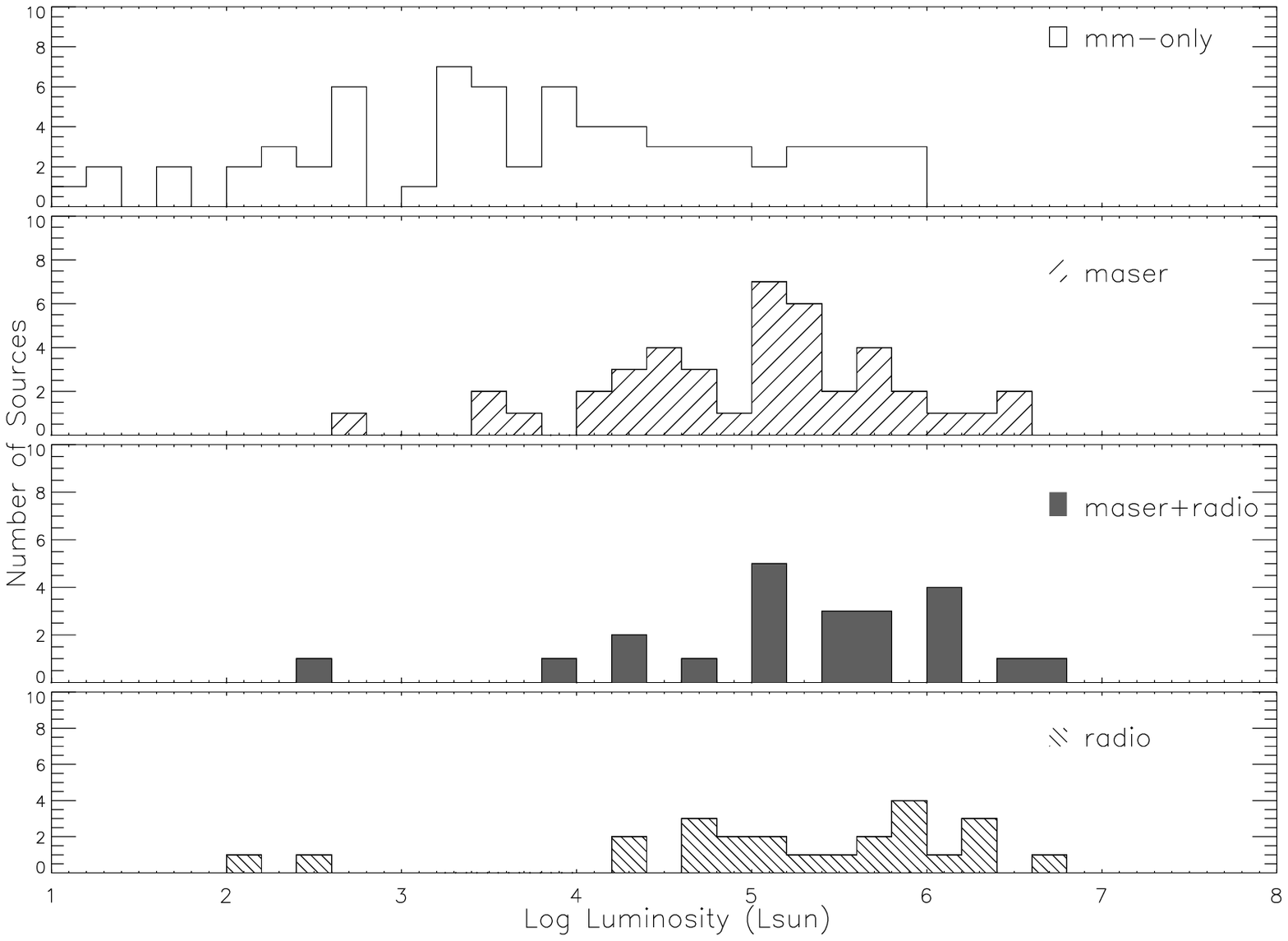} 
    \includegraphics[height=5.5cm, width=6.7cm]{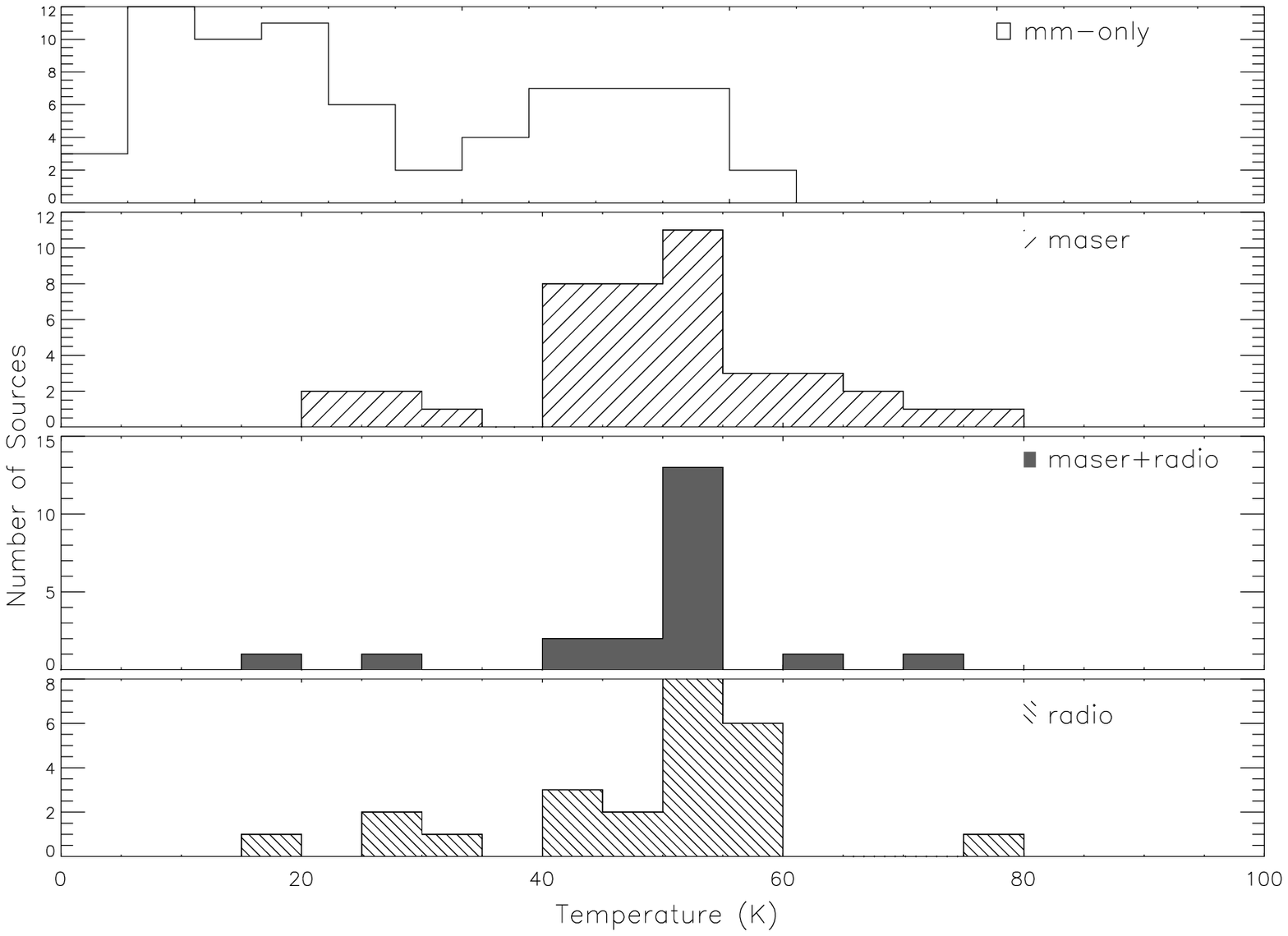}
    \caption{Histogram distribution of the luminosity ({\it left}) and temperature ({\it right}) for the four classes of source in the SIMBA sample.}
    \label{fig:hist}
  \end{center}
\end{figure}

\begin{figure}
\begin{center}
 \includegraphics[height=5.5cm, width=6.7cm]{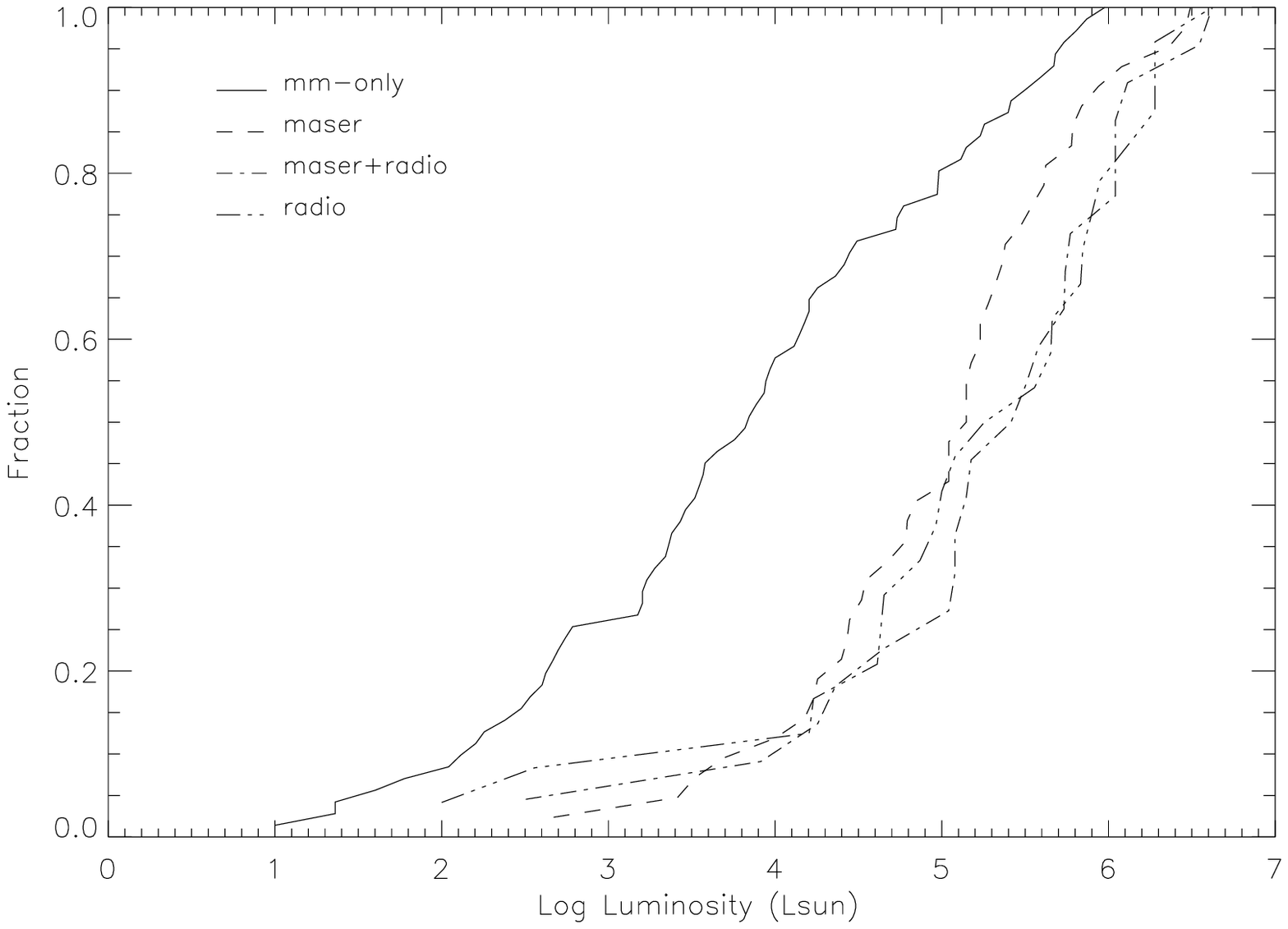}
 \includegraphics[height=5.5cm, width=6.7cm]{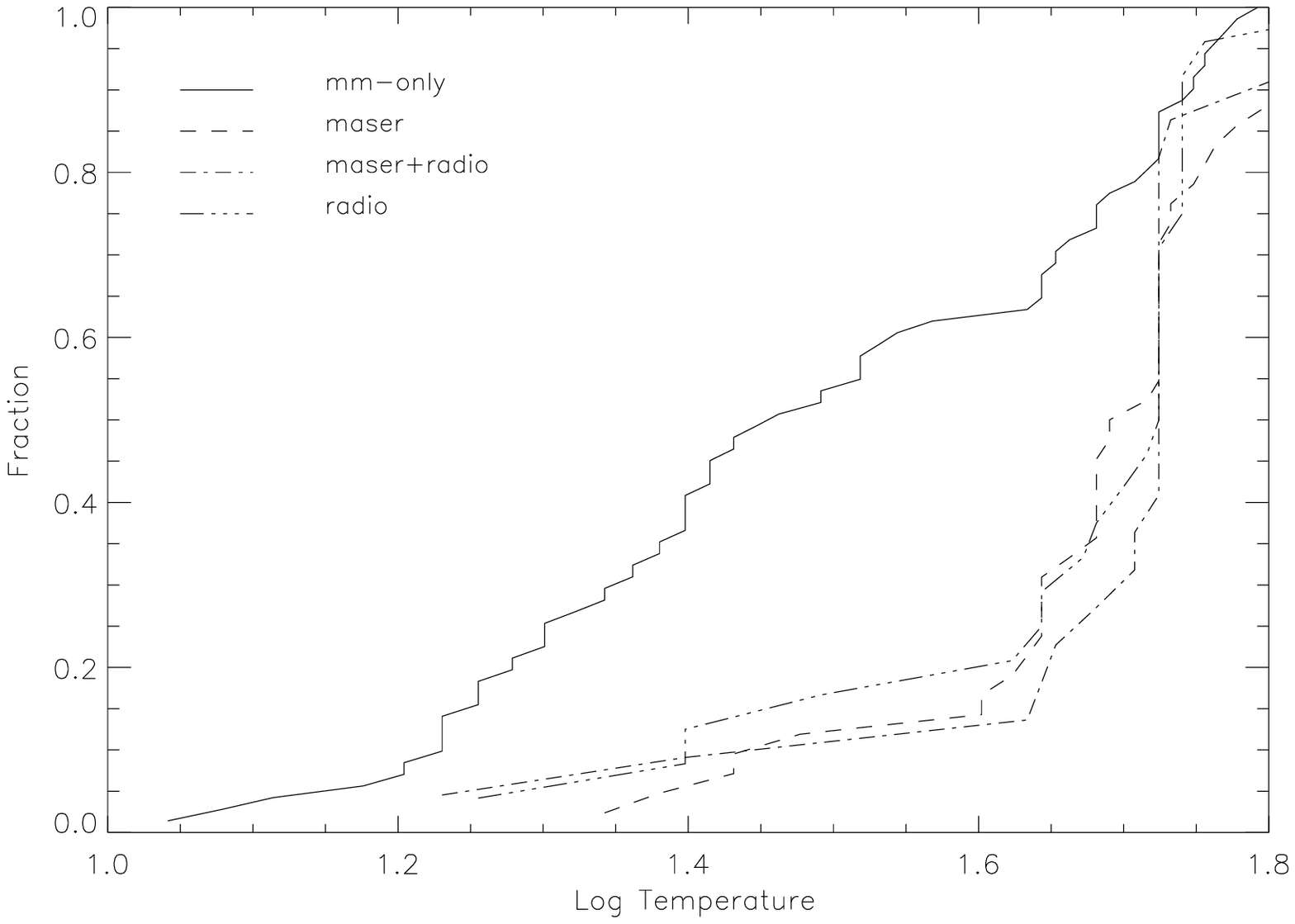}
  \caption{Cumulative distribution plots of the luminosity ({\it left}) and temperature {\it right}) for the four classes of source in the SIMBA sample.}
\label{fig:cumul}
\end{center}
\end{figure}

   We have compared each of these eight parameters with respect to the four classes of source found in the sample (see Paper I) through histogram and cumulative distribution plots (see figures \ref{fig:hist} and \ref{fig:cumul}) as well as Kolmogorov-Smirnov (KS) tests.

   Analysis revealed \citep{hill06t} the mm-only sample to be comparable in terms of mass and radii to sources harbouring a methanol maser and/or radio continuum source. Intriguingly, the mm-only sample are also cooler and less luminous (see Fig. \ref{fig:hist} and \ref{fig:cumul}) with smaller luminosity-to-mass ratios than these sources. As well as corroborating these results, the KS tests for the temperature, luminosity and L/M indicate that the mm-only cores are distinctly different from the methanol maser and/or radio continuum associated cores for these parameters. At least 45 per cent of the mm-only cores are also without mid-infrared \MSX\, emission.

   These results are consistent with cooler, younger examples of massive star formation, suggesting that the mm-only core is indicative of the very earliest stages of massive star formation, prior to the onset of methanol maser emission.

\section{A Bimodal Distribution for the mm-only Core?}\label{bi-mm}

   The mm-only sample was identified as comprising two populations which are distinguished by temperature (Fig. \ref{fig:hist}). For the following analysis, the mm-only sample was segregated into cool-mm and warm-mm components, creating five classes of sources in total.
  
\begin{figure}
\begin{center}
 \includegraphics[height=5.5cm, width=6.7cm]{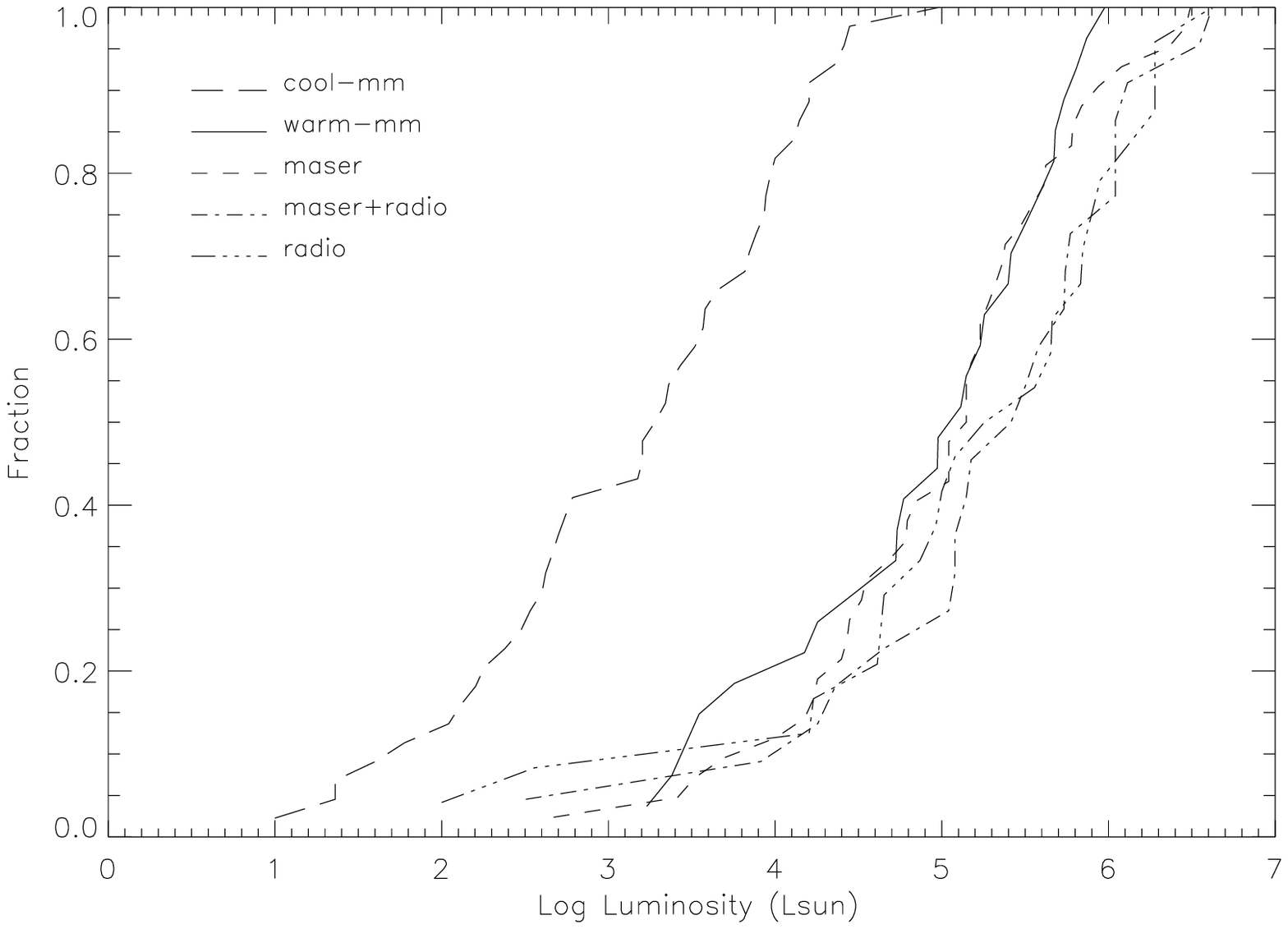}
 \includegraphics[height=5.5cm, width=6.7cm]{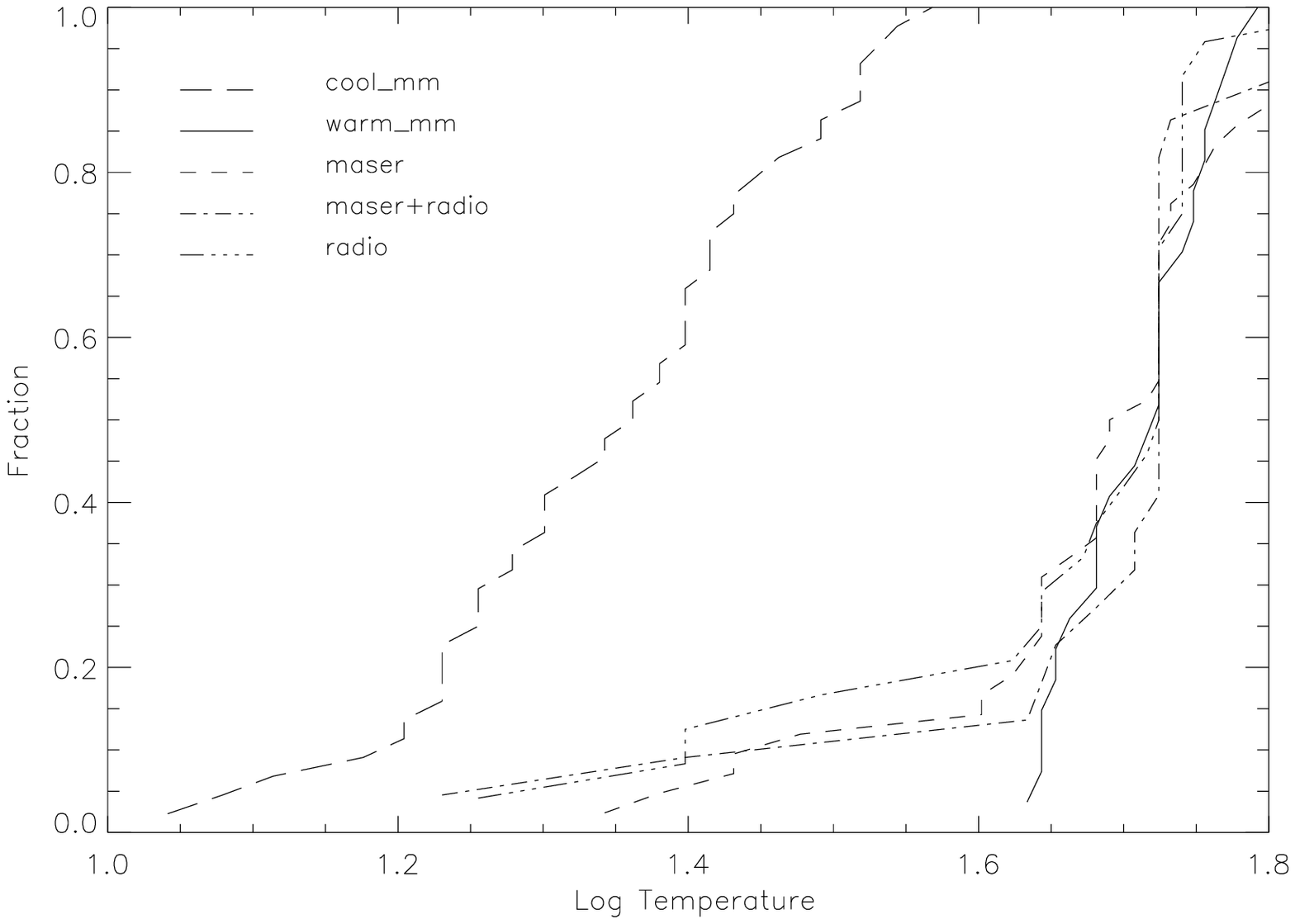}
  \caption{Cumulative distribution plots of the luminosity ({\it left}) and temperature ({\it right}) highlighting the two populations of mm-only core (cool-mm and warm-mm).}
\label{fig:bimm}
\end{center}
\end{figure}

   According to the cumulative distribution plots (cf. Fig. \ref{fig:bimm}) the cool-mm sources are less luminous with smaller L/M and radii, and higher densities (n$_{H_2}$ and $\Sigma$) than the warm-mm sources as well as those sources with a methanol maser and/or radio continuum source. The KS tests indicate that the cool-mm sources are not drawn from the same population as the warm-mm sources, nor the sources with radio continuum and/or methanol maser emission for the luminosity, luminosity-to-mass ratio, H$_2$ number density and the surface density. However no such distinction could be discerned for the mass.

   The warm-mm sources on the other hand display little distinction from the methanol maser and/or radio continuum sources for all of the parameters tested. These results led us to conclude that the cool-mm and warm-mm sources are two distinct populations of the mm-only core. The cool-mm sources are the least luminous sources in the sample, with lower luminosity-to-mass ratios as well as higher H$_2$ and surface densities. However they are of comparable mass and radii to the warm-mm and maser sources.

   We propose that the warm-mm sources are indicative of the earliest stages of massive star formation prior to the onset of methanol maser emission, whilst the cool-mm sources are examples of starless cores as described by \citet{vaz05}. High resolution observations of these cores, to search for embedded sources, are necessary in order to ascertain which of them are forming stars, and which will remain starless, and so test this hypothesis.

\section{Future Work}\label{concl}

  A caveat of this work is that the separation of the mm-only core into two distinct populations was concluded from a simple greybody fit to the SED, which enabled a temperature estimation in addition to the mass and luminosity of the dust cores. This distinction may be in error, a result of the limited number of data points available to fit the SEDs for these sources. 

   Ammonia (NH$_3$) is an excellent tracer of dense molecular gas. It can be used to probe the physical conditions of the molecular cloud in which cores are forming and can be used to determine physical conditions of the gas, such as temperature, density and optical depth \citep[cf.][]{longmore06}.  Intriguingly, Longmore et al. (this volume) report observations where the youngest star formation regions in their sample are devoid of methanol maser emission, traced solely by their NH$_3$ emission.

   We anticipate a direct correlation between future NH$_3$ observations of our sample, with the millimetre continuum emission detected by SIMBA, as per the results of \citet{pillai06}. Ammonia will therefore provide an independent determination of the temperature of our mm-only cores, with insights into their role in the formation of massive stars.

\begin{acknowledgments} 

The authors would like to thank S. Lumsden for his \MSX\, script, as well as C. Purcell and S. Longmore for assistance with converting the \MSX\, and \IRAS\, images. T.H. was supported by the UNSW Faculty Research Grant Program (FRGP) during her thesis.

\end{acknowledgments}

\end{document}